\documentclass{iopart}
\usepackage{iopams}
\usepackage{graphicx}
\usepackage{cite}

\newcommand{\proj}[1]{\langle #1 \rangle}

\begin{document}

\title{The Kustaanheimo-Stiefel transformation in geometric algebra}
\author{T Bartsch}
\address{Institut f\"ur Theoretische Physik 1, Universit\"at Stuttgart,
         70550 Stuttgart, Germany}
\ead{bartsch@theo1.physik.uni-stuttgart.de}

\begin{abstract}
The Kustaanheimo-Stiefel (KS) transformation maps the non-linear and
singular equations of motion of the three-dimensional Kepler problem to the
linear and regular equations of a four-dimensional harmonic oscillator. It
is used extensively in studies of the perturbed Kepler problem in celestial
mechanics and atomic physics. In contrast to the conventional matrix-based
approach, the formulation of the KS transformation in the language of
geometric Clifford algebra offers the advantages of a clearer geometrical
interpretation and greater computational simplicity. It is demonstrated
that the geometric algebra formalism can readily be used to derive a
Lagrangian and Hamiltonian description of the KS dynamics in arbitrary
static electromagnetic fields. For orbits starting at the Coulomb centre,
initial conditions are derived and a framework is set up that allows a
discussion of the stability of these orbits.
\end{abstract}


\section{Introduction}
\label{sec:Intro}

The Kepler problem belongs to the simplest systems of classical
mechanics. A detailed understanding of its properties is of equally
fundamental importance to celestial mechanics and to atomic physics.
In both areas it is of interest to describe, either by means of analytic
approximations or numerical computation, the impact of additional
non-Coulombic forces on the dynamics. In its original form, the equation of
motion
\begin{equation}
  \label{KeplerEq}
  \frac{\rmd^2 \bi x}{\rmd t^2} = -\frac{\bi x}{|\bi x|^3}
\end{equation}
is not well suited to this purpose because it is highly non-linear and
exhibits a singularity at the Coulomb centre where the force diverges. For
numerical studies of the dynamics it is mandatory to find a representation
of the equations of motion which avoids this singularity.

For the one-dimensional Kepler motion, it was already found by Euler
\cite{Euler1765} that the introduction of a square-root coordinate
$u=\sqrt{x}$ and a fictitious time $\tau$ defined by $\rmd t = x\,\rmd\tau$
reduces the Kepler equation of motion~(\ref{KeplerEq})
to the equation of motion of a one-dimensional harmonic oscillator
\begin{equation}
  \label{Dim1Reg}
  \frac{\rmd^2 u}{\rmd s^2} + 2 E u = 0 \;,
\end{equation}
where $E$ is the energy of the Kepler motion. Equation~(\ref{Dim1Reg}) is
not only void of singularities, it is also linear and thus forms a much
more convenient basis for analytic calculations.

Generalizing this approach, Levi-Civit\`a \cite{Levi56} regularized the
two-dimensional Kepler problem by combining the two spatial coordinates
into a complex number $x = x_1 + \rmi x_2$ and introducing a complex
square-root coordinate $u=\sqrt{x}$, which together with the
fictitious-time transformation $\rmd t = |x| \,\rmd\tau$ reduces the Kepler
problem to a two-dimensional harmonic oscillator.

Attempts to extend this regularization scheme to the three-dimensional
\mbox{Kepler} problem failed, until in 1964 Kustaanheimo and Stiefel
\cite{Kust64,KS65} proposed the introduction of four regularizing
coordinates instead of three and thereby achieved the reduction of the
three-dimensional Kepler problem to a four-dimensional harmonic
oscillator. This transformation, which is known as the Kustaanheimo-Stiefel
(KS) transformation, is discussed in detail in the monograph by Stiefel and
Scheifele \cite{Stiefel71}. Beyond its importance to celectial mechanics,
it has proven to be an essential tool for investigating the complicated
classical dynamics of the hydrogen atom in crossed electric and magnetic
fields \cite{Gourlay93,Milczewski97b,Sadovskii98}.

Customarily, the KS transformation is expressed in the language of matrix
algebra, which not only necessitates awkward computations of vector and
matrix components, but also lacks a transparent geometric interpretation.
An alternative formulation in terms of the geometric algebra
of Euclidean three-space was introduced by Hestenes~\cite{Hestenes90}. In
this formalism, the four KS coordinates are interpreted as the components
of a position spinor and are thus given a clear geometric
meaning. In addition, the formalism offers computational advantages
over the conventional matrix-based approach because it unites the four
coordinates into a single spinor.

In this paper I elaborate on the geometric algebra formulation of the KS
transformation and work out the details necessary for applications to
atomic physics, in particular semiclassical closed-orbit theory
\cite{Du88,Bogomolny89,Ellerbrock91,Main91,Bartsch02,Bartsch03b}. The
calculations will amply demonstrate the advantages of the geometric algebra
over the matrix-based formalism.  The terminology used is adapted to
applications in atomic physics. In particular, I refer to the attractive
centre as the nucleus and discuss the motion of an electron of unit mass
and negative unit charge under the influence of the nucleus and arbitrary
static external electromagnetic fields. Nevertheless, the results are of
much broader range. They apply equally to celestial mechanics or any other
field of physics governed by the equations of motion of a perturbed Kepler
problem.

In section~\ref{sec:KSequation}, the geometric algebra formulation of the
KS transformation is introduced and its relation to the matrix-based
approch is established.  The spinor equation of motion given by Hestenes
\cite{Hestenes90} is derived. In section~\ref{sec:KScan}, a Lagrangian and
Hamiltonian formulation of the KS spinor dynamics is derived. It cannot be
obtained by a simple change of variables because the KS transformation
introduces a non-physical fourth degree of freedom and a pseudotime
parameter. Nevertheless, it is demonstrated that the geometric algebra
formulation readily lends itself to an incorporation into the Lagrangian
and Hamiltonian formulations of dynamics. At the same time, the well-known
KS Hamiltonian \cite{Ellerbrock91,Main91,Gourlay93}, which is restricted to
homogeneous external electromagnetic fields, is generalized to arbitrary
static fields. Section~\ref{sec:Kepler} presents the explicit solution of
the spinor equation of motion in a pure Coulomb field and gives the
constants of motion in terms of the KS spinor. Section~\ref{sec:KSClosed}
discusses the KS description of orbits starting at the Coulomb centre and
returning to it. These orbits, referred to as ``closed orbits'', are of
particular importance because they play a central role in the semiclassical
interpretation of atomic photo-absorption spectra
\cite{Du88,Bogomolny89,Ellerbrock91,Main91}. They require a special
treatment because, although the spinor equation of motion is regular at the
nucleus, the KS transform is singular there. I will derive initial
conditions for orbits starting at the nucleus and present a basis of the
spinor space that is suitable for the investigation of the stability of
closed orbits.  A brief introduction into the properties of geometric
algebra needed here is given in \ref{app:GA}, where the notation used in
what follows is also explained. A more detailed exposition of the formalism
can be found in \cite{Hestenes90,Hestenes84,Lasenby93,Gull93}.

\section {The spinor equation of motion}
\label{sec:KSequation}

The KS-transformation in three dimensions can be found by representing an
arbitrary position in space not by its position vector $\bi x$, but by a
position spinor, i.e. the rotation-dilatation operator transforming a
fixed reference vector into the position vector $\bi x$. As explained in
\ref{app:GA}, a rotation-dilatation of the reference vector
$\bsigma_3$ is represented in the geometric algebra by an even
multivector $U$ according to
\begin{equation}
  \label{KStrafo}
  \bi x = \frac{1}{2}\, U \bsigma_3 U^\dagger \;.
\end{equation}
The factor 1/2 was introduced here to stay in touch with earlier
applications of the KS-transformation to atomic dynamics
\cite{Ellerbrock91,Main91}, although the present formulation of the theory
would suggest dropping it. It implies the normalization
\begin{equation}
  \label{KSnorm}
  U^\dagger U = U U^\dagger = 2\,r = 2\,|\bi x| \;.
\end{equation}
Up to normalization, the ansatz~(\ref{KStrafo}) reproduces the square-root
coordinates introduced by Euler and Levi-Civit\`a, respectively, if it is
applied to spaces of one or two dimensions.

Given a position vector $\bi x$, the choice of the spinor $U$ is not
unique. More precisely, the gauge transformation
\begin{equation}
  \label{KSgauge}
  U \mapsto U \rme^{-I_3 \alpha/2}
\end{equation}
with arbitrary real $\alpha$ does not alter $\bi x$, because the
additional exponential factor describes a rotation of the reference vector
$\bsigma_3$ around itself.  This consideration immediately clarifies
why a position spinor representation in three dimensions must introduce a
fourth degree of freedom. In lower dimensions, a rotation does not leave
any vector invariant, so that the spinor transformation does not possess a
gauge degree of freedom. It is also clear from~(\ref{KSgauge}) that all
fibres of the KS transformation except $U=0$ are circles in spinor space.

The inverse KS transformation can be found from equation (\ref{abRotor}) by
adapting the normalization to (\ref{KSnorm}). The position spinors
corresponding to a vector $\bi x$ are given by
\begin{equation}
  \label{invKS}
  U = \frac{r + \bi x \bsigma_3}{\sqrt{r+z}}\, \rme^{-I_3 \alpha/2}
\end{equation}
with arbitrary real $\alpha$.

In components, the spinor $U$ can be represented as $U=u_0 + I \bi u$ with
$\bi u=\sum_{k=1}^3 u_k \bsigma_k$. The transformation (\ref{KStrafo})
then decomposes into
\begin{eqnarray}
  \label{KScomp}
  \eqalign{
    x &= u_1 u_3 - u_0 u_2 \;, \\
    y &= u_1 u_0 + u_2 u_3 \;, \\
    z &= \frac{1}{2} (u_0^2-u_1^2-u_2^2+u_3^2) \;.
  }
\end{eqnarray}
Up to renumbering the components, this agrees with the conventions of
\cite{Ellerbrock91,Main91}.

To obtain an equation of motion for $U$, time derivatives of $U$ must be
calculated. Differentiating (\ref{KStrafo}) leads to
\begin{equation}
  \label{KSderiv}
  \dot{\bi x} = \frac 12 \dot U \bsigma_3 U^\dagger
              +\frac 12 U \bsigma_3 \dot U^\dagger
             = \proj{\dot U \bsigma_3 U^\dagger}_1 \;.
\end{equation}
Equation (\ref{KSderiv}) obviously cannot be solved for $\dot U$ because
the time derivative of the gauge parameter $\alpha$ in (\ref{KSgauge})
cannot be determined from the dynamics of the position vector. To arrive at
an equation of motion for $U$, I must therefore impose a constraint on
$\alpha$. This can be done in a convenient and geometrically appealing
fashion by requiring
\begin{equation}
  \label{KSconst}
  \proj{\dot U \bsigma_3 U^\dagger}_3 = 0 \;,
\end{equation}
which means that $\dot U$ is chosen such as not to contain a
component of rotation around the instantaneous position vector $\bi
x$. Under this constraint, (\ref{KSderiv}) yields
\begin{equation}
  \dot{\bi x} = \dot U \bsigma_3 U^\dagger
\end{equation}
and
\begin{equation}
  \label{KSderiv2}
  \dot U = \dot{\bi x}\, {U^\dagger}^{-1} \,\bsigma_3
         = \dot{\bi x}\, \frac{U}{2r} \,\bsigma_3 \;.
\end{equation}

As in the one- and two-dimensional cases, the regularization of the
three-dimensional Kepler motion requires the introduction of a
fictitious-time parameter $\tau$. It is defined by
\begin{equation}
  \label{KStauDef}
  \rmd t = 2r \, \rmd\tau
\end{equation}
Derivatives with respect to $\tau$ will be denoted with a prime.
Equation~(\ref{KSderiv2}) then yields
\begin{equation}
  \label{KSderivTau}
  U' = 2r \, \dot U = \dot{\bi x} \,U \,\bsigma_3 \;.
\end{equation}
For the second derivative of $U$, I obtain
\begin{eqnarray}
  \label{Udd}
  \eqalign{
    U'' &=    \left(\frac{d}{d\tau}\dot{\bi x}\right) \, U \bsigma_3 
            + \dot{\bi x} \, U' \,\bsigma_3 \\
        &= 2r \ddot{\bi x} \, U \bsigma_3 \,\frac{U^\dagger}{2r}\,U
            + {\dot{\bi x}^2} \, U \\
        &= 2 \left( \ddot{\bi x} \bi x + \frac 12 {\dot{\bi x}}^2
            \right) U \;.
  }
\end{eqnarray}
Together with Newton's equation of motion
\begin{equation}
  \label{NewtEq}
  \ddot{\bi x} = -\frac{\bi x}{r^3} + \bi f
\end{equation}
with an arbitrary non-Coulombic force $\bi f$, equation~(\ref{Udd}) yields
the spinor equation of motion in the form first given by Hestenes
\cite{Hestenes90}:
\begin{equation}
  \label{Ueq}
  U'' = 2\left( E_{\rm K} + \bi f \,\bi x \right) U \;,
\end{equation}
where the Kepler energy
\begin{equation}
  \label{Ek}
  E_{\rm K} = \frac 12 \dot{\bi x}^2 - \frac 1r
\end{equation}
denotes the sum of the kinetic and Coulombic potential energies.

In the special case of pure Kepler motion, i.e.~$\bi f=0$, the Kepler
energy $E_{\rm K}$ is equal to the total energy $E$ and is conserved. In
this case, (\ref{Ueq}) reduces to the linear equation of motion
\begin{equation}
  \label{UKepler}
  U'' = 2 E \,U \;.
\end{equation}
If $E<0$, this is the equation of motion of a  four-dimensional isotropic
harmonic oscillator with frequency $\omega=\sqrt{-2E}$ with respect to
$\tau$.

If additional forces $\bi f$ are present, the Kepler energy is not
conserved in general, so that the work done by the external forces must be
taken into account \cite{KS65}. This can easily be achieved if the external
forces are generated by static electromagnetic fields, because the work
done by a magnetic field $\bi B$ is zero, whereas an electric field $\bi
F=-\nabla V$ can be derived from a potential $V(\bi x)$. In this case, the
total energy $E=E_{\rm K}-V$ is conserved, so that the equation of motion
reads
\begin{equation}
  \label{UField}
  U'' = 2 \left(E+V(\bi x) + \bi f \,\bi x\right) U
\end{equation}
with $\bi f = -\bi F - \dot{\bi x}\times\bi B$.

It finally remains to verify that the equation of motion (\ref{Ueq}) is
consistent with the constraint (\ref{KSconst}). To prove this, first note
that
\begin{equation}
  \label{xiDef}
  \xi = \proj{U' \bsigma_3 U^\dagger}_3
\end{equation}
is a constant of motion for any external forces $\bi f$ \cite{KS65},
because by~(\ref{Ueq})
\begin{eqnarray}
  \label{XiDeriv}
  \eqalign{
    \frac{d\xi}{d\tau} &=
      \proj{U''\,\bsigma_3\,U^\dagger}_3 +
        \underbrace{\proj{U'\,\bsigma_3\,{U'}^\dagger}_3}
                  _{\textrm{=0 by~(\ref{UVector})}}\\
    &= \proj{\left(E_{\rm K}+\bi f\,\bi x\right) \bi x}_3 \\
    &= \proj{E_{\rm K}\bi x + r^2 \bi f}_3 \\
    &= 0 \;.
  }
\end{eqnarray}
Therefore, if the initial conditions are chosen so that $\xi =0$ at
$\tau=0$, equation~(\ref{XiDeriv}) guarantees $\proj{\dot U \,\bsigma_3
\,U^\dagger}_3 = 2r\,\xi = 0$ at all times.

\section{Canonical formalism}
\label{sec:KScan}

In classical investigations of atoms in external fields, the
Hamiltonian nature of the dynamics plays a central role. It is therefore
essential to show how the spinor equation of motion found in the previous
section can be derived in the context of a Lagrangian or Hamiltonian
formalism.  
In the matrix theory of the KS transformation, a Hamiltonian formulation
is well known and widely applied in the literature
\cite{Ellerbrock91,Main91}. Due to the introduction of an
additional degree of freedom and a fictitious-time parameter, it cannot be
found by a straightforward change of variables. In this section it will be
shown that the application of geometric algebra allows an easy and
general derivation of the Hamiltonian. At the same time, the Hamiltonian
formalism will be generalized to arbitrary inhomogeneous static external
fields.

\subsection{Fictitious-time transformations}
\label{ssec:FictTime}

Elementary expositions of Lagrangian and Hamiltonian dynamics usually treat
the time $t$ as the externally prescribed independent variable
fundamentally different from the spatial coordinates, velocities, and
momenta. The formalisms are then shown to be invariant under point
transformations or canonical transformations, respectively, which may be
time-dependent, but may not transform the time variable. However, both the
Lagrangian and the Hamiltonian formalisms can be reformulated in such a way
that it is possible to introduce an arbitrary orbital parameter $\tau$ and
to treat the physical time $t$ as an additional coordinate on the same
footing as the spatial coordinates. This formalism is
discussed in its full generality by Dirac \cite{Dirac33, Dirac50}.
For the special case of autonomous Lagrangian dynamics and the simple form
of the fictitious-time transformation used above, the full flexibility of
Dirac's homogeneous formalism is not needed. Instead, the modifications
needed to achieve the fictitious-time transformation can be derived in a
straightforward manner from Hamilton's variational principle.

The Lagrangian equations of motion can be derived from the action
functional
\begin{equation}
  \label{actDef}
  S = \int_{t_1}^{t_2} \rmd t \, L(q(t), \dot q(t))
\end{equation}
by requiring that for the classical paths the variation of $S$ with respect
to the path $q(t)$ vanishes if the variation is performed with the initial
and final times $t_1$ and $t_2$ and the coordinates $q(t_1)$ and $q(t_2)$
kept fixed.
If a fictitious-time parameter $\tau$ is introduced by the prescription
\begin{equation}
  \label{genTauDef}
  \rmd t = f(q,\dot q)\, \rmd \tau
\end{equation}
with an arbitrary function $f$, it is tempting to rewrite the action
functional as
\begin{equation}
  \label{actRewrite}
  S = \int_{\tau_1}^{\tau_2} \rmd\tau \, f(q, \dot q) L(q,\dot q)
\end{equation}
and regard
\begin{equation}
  \label{tildeL}
  \tilde L = f(q, \dot q)\, L (q, \dot q) = \frac{\rmd t}{\rmd\tau}\, L
\end{equation}
as the Lagrangian describing the dynamics with respect to $\tau$. However,
this simple procedure is incorrect in general, because to derive the
Lagrangian equations with respect to $\tilde L$ from (\ref{actRewrite}),
the variation of $S$ has to be performed with the initial and final
fictitious times $\tau_1$ and $\tau_2$ kept fixed, and due
to~(\ref{genTauDef}) a variation of the path will alter the relation
between $t$ and $\tau$, so that the initial and final physical times $t_1$
and $t_2$ will vary.

To establish the true relation between (\ref{actDef}) and
(\ref{actRewrite}), I calculate the variation of (\ref{actRewrite}) taking
the variation of $t$ into account, i.e. $q$ and $t$ are varied according
to
\begin{eqnarray}
  \label{varDef}
  \eqalign{
    q(\tau) &\mapsto q(\tau)+\delta q(\tau) \;, \\
    t(\tau) &\mapsto t(\tau)+\delta t(\tau) \;,
  }
\end{eqnarray}
subject to the boundary conditions
\begin{equation}
  \label{boundCond}
   \delta q(\tau_1) = \delta q(\tau_2) = 0
\end{equation}
and with $\tau_1$ and $\tau_2$ kept fixed. Under this variation,
\begin{eqnarray}
  \eqalign{
    \frac{\rmd\tau}{\rmd t} \mapsto \, \frac{\rmd\tau}{\rmd(t+\delta t)} 
	&= \frac{1}{\frac{\rmd t}{\rmd\tau}+\frac{\rmd\,\delta t}{\rmd\tau}} \\
        &= \frac{\rmd\tau}{\rmd t}
              \left(1-\frac{\rmd\tau}{\rmd t}\,
                      \frac{\rmd\,\delta t}{\rmd\tau}\right)
    \;,
  }
\end{eqnarray}
so that
\begin{equation}
  \delta\,\frac{\rmd\tau}{\rmd t} = -\left(\frac{\rmd\tau}{\rmd t}\right)^2
       \frac{\rmd\,\delta t}{\rmd\tau}
\end{equation}
and hence
\begin{eqnarray}
  \eqalign{
    \delta\dot q
       &= \delta \left(\frac{\rmd q}{\rmd\tau}\,
                       \frac{\rmd\tau}{\rmd t}\right) \\
       &= \delta\left(\frac{\rmd q}{\rmd\tau}\right) 
                     \cdot \frac{\rmd\tau}{\rmd t} +
                \frac{\rmd q}{\rmd\tau} \cdot 
                     \delta\left(\frac{\rmd\tau}{\rmd t}\right) \\
       &= \frac{\rmd\,\delta q}{\rmd t} 
             - \dot q \,\frac{\rmd\tau}{\rmd t}\,
                        \frac{\rmd\,\delta t}{\rmd\tau} \;.
  }
\end{eqnarray}
The variation of (\ref{actRewrite}) then reads
\begin{eqnarray}
  \label{actVar}
  \eqalign{
    \delta S 
      &= \int \rmd\tau \left[ \delta\left(\frac{\rmd t}{\rmd\tau}\right)\,L +
         \frac{\rmd t}{\rmd\tau} \left(\frac{\partial L}{\partial q}\,\delta q 
                 + \frac{\partial L}{\partial \dot q}\,\delta\dot q \right)
         \right] \\
      &= \int \rmd\tau\,\frac{\rmd\,\delta t}{\rmd\tau}
            \left(L-\dot q\,\frac{\partial L}{\partial\dot q}\right) +
         \int \rmd t\,\left(\frac{\partial L}{\partial q}\,\delta q +
                    \frac{\partial L}{\partial\dot q}\,
                    \frac{\rmd\,\delta q}{\rmd t}
                  \right) \\
      &= -\int \rmd\tau\,\frac{\rmd\,\delta t}{\rmd\tau}\,H +
         \int \rmd t\,\left(\frac{\partial L}{\partial q} - 
                        \frac{\rmd}{\rmd t}\,\frac{\partial L}{\partial\dot q}
                  \right) \delta q \;,
  }
\end{eqnarray}
where the customary partial integration was performed, the boundary
conditions (\ref{boundCond}) were used and the Hamiltonian
\begin{equation}
  \label{HamDef}
  H = \dot q\,\frac{\partial L}{\partial\dot q} - L
\end{equation}
was introduced.

If only the second integral in the last line of (\ref{actVar}) were present,
it would yield the correct equations of motion. Thus, the action
functionals (\ref{actDef}) and (\ref{actRewrite}) are equivalent if the
Hamiltonian $H$ vanishes. For autonomous systems, $H$ is a constant of
motion equal to the energy $E$. If the Lagrangian $L$ is replaced with~$L+E$,
with $E$ regarded as a constant, the equations of motion derived from $L$
are unchanged, but the Hamiltonian (\ref{HamDef}) changes to $H-E=0$. Thus,
the dynamics of trajectories with energy~$E$ with respect to the
fictitious-time parameter~$\tau$ is described by the Lagrangian
\begin{equation}
  \label{tauLag}
  {\cal L} = \frac{\rmd t}{\rmd\tau}\,(L+E) = f(q,\dot q)\,(L+E)\;.
\end{equation}
This Lagrangian has to be written as a function of the coordinates $q$ and
the fictitious-time velocities $q'$. If the function $f$
is independent of the velocities, the canonical momenta are invariant under
the fictitious-time transformation, because $q'=f(q)\,\dot q$ and
\begin{equation}
  \frac{\partial\cal L}{\partial q'} = 
   f(q)\,\frac{\rmd\dot q}{\rmd q'}\,\frac{\partial L}{\partial\dot q} =
   \frac{\partial L}{\partial\dot q} \;.
\end{equation}

From the time-transformed Lagrangian (\ref{tauLag}), the transformed
Hamiltonian
\begin{equation}
  \label{tauHam}
  {\cal H} = q'\,\frac{\partial\cal L}{\partial q'} - {\cal L}
           = f(q,\dot q)\,(H-E)
\end{equation}
is obtained by the usual Legendre transformation. It must be written as a
function of the coordinates and momenta. In some cases the passage from the
Lagrangian to the Hamiltonian description of the dynamics is impossible
because the relation $p=\partial{\cal L}(q,q')/\partial q'$ cannot be
solved for $q'$. In these cases, the Hamiltonian (\ref{tauHam}) can be
shown to describe the fictitious-time dynamics by a discussion of the
modified Hamilton's principle analogous to the derivation of the
Lagrangian~$\cal L$ above.

\subsection{Lagrangian description}
\label{ssec:KSLag}

The dynamics of an atomic electron under the combined influences of the
nuclear Coulomb potential, an additional scalar potential $V(\bi x)$ and a
magnetic field represented by a vector potential $\bi A(\bi x)$ is
described by the Lagrangian
\begin{equation}
  \label{origLag}
  L = \frac{\dot{\bi x}^2}{2}+\frac{1}{r}+V(\bi x)-\bi A\cdot\dot{\bi x}\;.
\end{equation}
This Lagrangian must be transformed to a Lagrangian $\cal L$ describing the
fictitious-time dynamics of the position spinor $U$. With $f(q) = 2r =
U^\dagger U$ by~(\ref{KStauDef}), the fictitious-time
Lagrangian~(\ref{tauLag}) reads
\begin{equation}
  \label{tauLagNoConst}
  \fl
  {\cal L} = \frac 14\,{U'}^\dagger U' 
            +\frac{1}{8r}\,\left\langle\left(U'\bsigma_3 U^\dagger\right)^2
                           \right\rangle
            + E\,U^\dagger U 
            + U^\dagger U \,V(\bi x)
            - \proj{\bi A(\bi x)\,U'\bsigma_3 U^\dagger} + 2
\end{equation}
with $\bi x=\frac 12  U\bsigma_3 U^\dagger$.
If the constraint~(\ref{KSconst}) is used, $\cal L$ simplifies to
\begin{equation}
  \label{tauLagConst}
  \fl
  {\cal L} = \frac 12\,{U'}^\dagger U' + E\,U^\dagger U 
            + U^\dagger U \,V(\bi x)
            - \proj{\bi A(\bi x)\, U'\bsigma_3 U^\dagger} + 2
\end{equation}
Both forms of the Lagrangian yield the same ``on-shell''
dynamics for trajectories satisfying~(\ref{KSconst}).
Note that only the kinetic term is influenced by the
constraint, whereas potential and vector potential terms are not.

The momentum conjugate to $U$ is given by
\begin{equation}
  \label{PNoConst}
  P = \partial_{U'}{\cal L} = \frac{1}{2r}\,\bsigma_3 U^\dagger
     \proj{U'\bsigma_3 U^\dagger}_1 - \bsigma_3 U^\dagger \bi A \;,
\end{equation}
which simplifies to
\begin{equation}
  \label{PConst}
  P = {U'}^\dagger - \bsigma_3 U^\dagger \bi A
\end{equation}
if~(\ref{KSconst}) is applied.

As the spinor equation of motion (\ref{Ueq}) is valid under the constraint
(\ref{KSconst}) only, the Lagrangian $\cal L$ provides a suitable
description of the dynamics if it reproduces~(\ref{Ueq}) for trajectories
satisfying~(\ref{KSconst}). The simplified Lagrangian~(\ref{tauLagConst})
can therefore be used. When equations of motion are derived
from~(\ref{tauLagConst}), the constraint~(\ref{KSconst}) must be taken into
account by a Lagrangian multiplier. I will now show, however, that the
unconstrained equation of motion
\cite{Lasenby93}
\begin{equation}
  \label{genLagEq}
  \frac{d}{d\tau}\,\partial_{U'}{\cal L} - \partial_U {\cal L} = 0
\end{equation}
derived from (\ref{tauLagConst}) reproduces (\ref{Ueq}) without the
constraint being explicitly dealt with, i.e.~the Lagrangian
multiplier to be introduced turns out to vanish identically. I
therefore ignore it from the outset.

For the case of vanishing external potentials, (\ref{genLagEq}) can easily
be seen to yield
\begin{equation}
  \label{KeplerLagEq}
  {U''}^\dagger - 2 E U^\dagger = 0 \;,
\end{equation}
which is the reversion of (\ref{UKepler}). For the terms containing the
potentials, the calculation of~(\ref{genLagEq}) is still straightforward,
but requires a more intimate familiarity with the properties of the
multivector derivative. I will therefore present the calculation in detail.

The contribution of the scalar potential term
\begin{equation}
  \label{VDef}
  {\cal V} = U^\dagger U \, V(\bi x(U))
\end{equation}
with $\bi x(U) = \frac 12 U \bsigma_3 U^\dagger$ reads,
by~(\ref{prodMult}),
\begin{equation}
  \label{VU1}
  \partial_U {\cal V} = 2 U^\dagger V(\bi x) +
                        U^\dagger U \, \partial_U V(\bi x(U)) \;.
\end{equation}
The chain rule (\ref{chainDirect}) then yields for any even multivector $M$
\begin{equation}
  M\ast\partial_U V(\bi x(U)) 
    = \left(M\ast\partial_U\,\bi x(U)\right) \ast \partial_{\bi x} V
\end{equation}
with
\begin{eqnarray}
  \label{xDiff}
  \eqalign{
    M\ast\partial_U\,\bi x(U)
     &= (M\ast \partial_U) \frac 12 \,U \bsigma_3 U^\dagger \\
     &= \frac 12 M \bsigma_3 U^\dagger + 
        \frac 12\, U \bsigma_3 M^\dagger \\
     &= \proj{M \bsigma_3 U^\dagger}_1 \;.
  }
\end{eqnarray}
In the absence of a magnetic field the external force is $\bi f =
\partial_{\bi x} V$, so that
\begin{equation}
   M\ast\partial_U V(\bi x(U))
    = \proj{\proj{M \bsigma_3 U^\dagger}_1 \bi f}
    = \proj{M \bsigma_3 U^\dagger \bi f} \;.
\end{equation}
Thus, 
\begin{equation}
  \partial_U V = \partial_M \left(M\ast\partial_U V\right) 
               = \bsigma_3 U^\dagger \bi f \;,
\end{equation}
and finally
\begin{equation}
  \label{VU}
  \partial_U {\cal V} = 2 U^\dagger V(\bi x) +
       2 U^\dagger \bi x \bi f \;.
\end{equation}
This is the reversion of the scalar-potential terms
in~(\ref{Ueq}). Therefore, (\ref{VU}) together with~(\ref{KeplerLagEq})
indeed yields the correct equation of motion.

To evaluate the contribution of the vector potential term
\begin{equation}
  \label{ADef}
  {\cal A} = \proj{\bi A(\bi x) U' \bsigma_3 U^\dagger}
\end{equation}
to (\ref{genLagEq}), first note that
\begin{eqnarray}
  \label{ACont1}
  \eqalign{
    \frac{d}{d\tau}\,\partial_{U'} {\cal A}
   &= \frac{d}{d\tau}\,
         \left(\bsigma_3 U^\dagger \bi A(\bi x)\right) \\
   &= \bsigma_3 {U'}^\dagger \bi A(\bi x) +
      \bsigma_3 U^\dagger (\bi x'\cdot\partial_{\bi x}) \bi A(\bi x)
   \;.
  }
\end{eqnarray}
By Leibniz' rule and (\ref{KSderiv}),
\begin{eqnarray}
  \label{ACont2}
  \eqalign{
    \partial_U {\cal A}
   &= \stackrel{*}{\partial}_U\proj{\bi A\, U'\bsigma_3 
	                           \stackrel{*}{U^\dagger}}
     +\stackrel{*}{\partial}_U\proj{\stackrel{*}{\bi A}\,
                                   U'\bsigma_3 U^\dagger} \\
   &= \bsigma_3 {U'}^\dagger \bi A + 
      \partial_U \proj{\bi A(\bi x(U))\, \bi x'} \;.
  }
\end{eqnarray}
The second term on the right-hand side can be evaluated by first
calculating directional derivatives. For an arbitrary even $M$,
(\ref{xDiff}) and the chain rule~(\ref{chainDirect}) yield
\begin{eqnarray}
  \eqalign{
    (M\ast\partial_U) \proj{\bi A(\bi x(U))\, \bi x'}
   &= \left(M\ast\partial_U \bi A\right) \ast\partial_{\bi A}
      \proj{\bi A \bi x'} \\
   &= \proj{\left(M\ast\partial_U \bi A(\bi x(U))\right) \bi x'} \\
   &= \proj{\left(M\ast\partial_U \bi x\right)\ast\partial_{\bi x} \,
            \bi A \bi x'} \\
   &=
      \proj{\proj{\proj{M\bsigma_3 U^\dagger}_1 \partial_{\bi x}}
            \bi A \bi x'} \\
   &= \proj{M\bsigma_3 U^\dagger\,\partial_{\bi x}} \proj{\bi A \bi x'}
   \;,
  }
\end{eqnarray}
so that
\begin{eqnarray}
  \label{ACont3}
  \eqalign{
    \partial_U  \proj{\bi A(\bi x(U)) \,\bi x'}
   &= \partial_M (M\ast\partial_U) \proj{\bi A(\bi x(U)) \,\bi x'} \\
   &= \bsigma_3 U^\dagger\, \partial_{\bi x} (\bi A\cdot\bi x') \;.
  }
\end{eqnarray}
Equations~(\ref{ACont1}), (\ref{ACont2}) and (\ref{ACont3}) combine to
\begin{eqnarray}
  \label{ACont}
  \eqalign{
    \frac{d}{d\tau}\,\partial_{U'}{\cal A} - \partial_U {\cal A}
   &=  \bsigma_3 U^\dagger\left((\bi x'\cdot\partial_{\bi x}) \bi A
                                  -\partial_{\bi x}(\bi A\cdot\bi x')
                             \right) \\
   &= -\bsigma_3 U^\dagger\left(\bi x'\times(\partial_{\bi x} \times
                                    \bi A)\right) \\
   &= -2\,U^\dagger \bi x (\dot{\bi x}\times\bi B) \;,
  }
\end{eqnarray}
which is the reversion of the magnetic-field contribution to (\ref{Ueq}).

Note that the derivation given here is valid for arbitrary external
potentials $V$ and $\bi A$, whereas conventional treatments restrict
themselves to the special case of homogeneous external fields. It can
obviously be further generalized to include conservative forces of
non-electromagnetic origin. Also note that the geometric algebra formalism
allows one to do the calculations in a straightforward manner without
having to resort to component decompositions of any of the vectorial or
spinorial quantities involved.

\subsection{Hamiltonian description}
\label{ssec:KSHam}

The transition from a Lagrangian to a Hamiltonian description of the
dynamics leads from the Lagrangian (\ref{tauLagNoConst}) or
(\ref{tauLagConst}), depending on whether or not the
constraint~(\ref{KSconst}) is applied, to the Hamiltonian
\begin{equation}
  {\cal H} = (U'\ast\partial_{U'}){\cal L} - {\cal L} \;,
\end{equation}
in which the velocity $U'$ has to be expressed in terms of the momentum
$P$. The transformation requires that the relation (\ref{PNoConst}) or
(\ref{PConst}) between velocity and momentum can be solved for the
velocity, which is impossible in the case of (\ref{PNoConst}). Thus, the
constraint (\ref{KSconst}) is not only needed to obtain an unambiguous
equation of motion for $U$, but also serves as a condition for a Hamiltonian
description of the spinor dynamics to exist. If it is imposed and an
inessential constant of 2 is added, the Hamiltonian reads
\begin{equation}
  \label{genTauHam}
  {\cal H} = \frac 12\left(P^\dagger + \bi A\, U \bsigma_3\right)
                     \left(P + \bsigma_3 U^\dagger \bi A\right)
             - E U^\dagger U - U^\dagger U \,V(\bi x)
           = 2 \;.
\end{equation}
Because it is time-independent, the Hamiltonian (\ref{genTauHam}) is a
constant of motion. To describe the physical dynamics, its value must be
chosen to be 2, whereas the physical energy $E$ appears as a parameter
in $\cal H$.

The equations of motion derived from (\ref{genTauHam}) read
\begin{eqnarray}
  \label{genHamEq}
  \eqalign{
    U' &= \phantom{-}\partial_P{\cal H} =
         P^\dagger + \bi A\,U \bsigma_3 \;,\\
    P' &= -\partial_U{\cal H} \\
       &= -2E\,U^\dagger - 2 U^\dagger V(\bi x) 
         - 2 U^\dagger\bi x\,\partial_{\bi x}V \\
       &\phantom{=\ }
         - \bsigma_3 \left(P+\bsigma_3 U^\dagger\bi A\right)\bi A
         - \bsigma_3U^\dagger\,
           \stackrel{*}{\partial}_{\bi x}\proj{\stackrel{*}{\bi A}\,\bi x'}
       \;,
  }
\end{eqnarray}
where
\begin{equation}
  \label{xPrHam}
  \bi x' = U\bsigma_3U'^\dagger
    = U\bsigma_3P + U^\dagger U\,\bi A
\end{equation}
was used. In terms of coordinates and momenta, the constraint
(\ref{KSconst}) reads
\begin{equation}
  \label{ConstHam}
  \proj{U \bsigma_3 P}_3 = 0 \;.
\end{equation}
Equation~(\ref{ConstHam}) is equivalent to~(\ref{KSconst}) both in the
presence and in the absence of a magnetic field.
Taken together, (\ref{genHamEq}) and (\ref{xPrHam}) lead back to the
equation of motion (\ref{Ueq}).

Finally, let me mention an important subtlety regarding the component
decomposition of the spinor equation. If, according to (\ref{KScomp}), $U$
is represented as $U=u_0+I\bi u$ with $\bi u=\sum_{k=1}^3
u_k\bsigma_k$ and $p_k$ denotes the momentum component conjugate to
$u_k$, the spinor momentum is $P=p_0-I\bi p$ with $\bi p=\sum_{k=1}^3
p_k\bsigma_k$. The negative sign is necessary because in the spinor
formulation the bivector $I_kp_k$ is conjugate to $I_ku_k$. Dropping
the bivector factors $I_k$ leads to the stated result.

\section{The Kepler problem}
\label{sec:Kepler}

The unperturbed Kepler motion is described by the Hamiltonian
\begin{equation}
  \label{KepHam}
  {\cal H} = \frac 12\,P^\dagger P - E\,U^\dagger U = 2 \;.
\end{equation}
The equations of motion derived from~(\ref{KepHam}) are
\begin{equation}
  \label{KepHamEq}
  U'=P^\dagger\;, \qquad P'=2E U^\dagger
\end{equation}
or, as in~(\ref{UKepler}),
\begin{equation*}
  U''=2 E U \;.
\end{equation*}
If $E<0$, this is the equation of motion of an isotropic four-dimensional
harmonic oscillator, whose general solution reads
\begin{equation}
  \label{KeplerSol}
   U=A\cos(\sqrt{-2E}\,\tau)+B\sin(\sqrt{-2E}\,\tau)
\end{equation}
with two constant even multivectors $A$ and $B$ bound, by (\ref{KSconst})
and (\ref{KepHam}), to satisfy
\begin{equation}
  \proj{A\bsigma_3 B^\dagger}_3 = 0
\end{equation}
and 
\begin{equation}
  A^\dagger A+B^\dagger B = -\frac2E \;.
\end{equation}

In the case of the pure Kepler motion, the angular momentum vector $\bi L$
and the Lenz vector $\boldsymbol\epsilon$ are conserved. Together, they
uniquely specify an orbit \cite{Hestenes90}. I will now derive the
KS-transformed expressions for these constants of motion. Throughout, the
validity of the constraint (\ref{KSconst}) will be assumed.

The angular momentum vector is given by
\begin{equation}
  \label{LVecDef}
  \bi L = \bi x \times \dot{\bi x} = -I\proj{\bi x \dot{\bi x}}_2 \;.
\end{equation}
Within the geometric algebra, it is more convenient to introduce the
angular momentum bivector
\begin{equation}
  \label{lBivDef}
  l = I \bi L = \proj{\bi x \dot{\bi x}}_2 \;,
\end{equation}
which specifies the orbital plane instead of the direction perpendicular to
it. By~(\ref{KStrafo}), (\ref{KSderiv}) and (\ref{KSconst}),
\begin{eqnarray}
  \label{lKS}
  \eqalign{
    l &= \proj{\frac 12\,U\bsigma_3U^\dagger 
               \, U\bsigma_3\dot U^\dagger}_2 \\
      &= \frac 12\proj{U{U'}^\dagger}_2 \\
      &= \frac 12 \proj{UP}_2 \;.
  }
\end{eqnarray}

That $l$ is conserved can be verified by a straightforward
differentiation. Alternatively, it can be checked that the
Poisson bracket vanishes,
\begin{eqnarray}
  \label{lPoisson}
  \eqalign{
    \left\{l,{\cal H}\right\}
   &= \left(\partial_P{\cal H}\right)\ast\partial_U l
     -\left(\partial_U{\cal H}\right)\ast\partial_P l \\
   &= P^\dagger\ast\partial_U l + 2E\,U^\dagger\ast\partial_P l \\
   &= \frac 12 \proj{P^\dagger P}_2 + E \proj{U^\dagger U}_2 \\
   &= 0 \;.
  }
\end{eqnarray}
Note how the Poisson bracket formalism extends not only to multivector
coordinates $U$ and $P$, but also to non-scalar arguments.

The Lenz vector is given by
\begin{eqnarray}
  \label{epsDef}
  \eqalign{
    \boldsymbol\epsilon &= l\dot{\bi x} - \frac{\bi x}{r} \\
      &= l\,P^\dagger\bsigma_3U^{-1} - U\bsigma_3U^{-1} \;.
  }
\end{eqnarray}
To calculate the Poisson bracket $\left\{\boldsymbol\epsilon,{\cal
H}\right\}$, use $\left\{l,{\cal H}\right\}=0$ and
\begin{equation}
  P^\dagger\ast\partial_U U^{-1} = -U^{-1}P^\dagger U^{-1}
\end{equation}
to find
\begin{eqnarray}
  \label{epsPoisson}
  \eqalign{
    \left\{\boldsymbol\epsilon,{\cal H}\right\} 
   &= l\left\{P^\dagger\bsigma_3U^{-1}, {\cal H}\right\}
      - \left\{U\bsigma_3U^{-1},{\cal H}\right\} \\
   &= l \left(P^\dagger\ast\partial_U\left(P^\dagger\bsigma_3U^{-1}\right)
             +2E\,U^\dagger\ast\partial_P
                         \left(P^\dagger\bsigma_3U^{-1}\right)
        \right) \\
      &\quad- P^\dagger\ast\partial_U\left(U\bsigma_3U^{-1}\right) \\
   &= l\left(-P^\dagger\bsigma_3U^{-1}P^\dagger U^{-1}
             +2EU\bsigma_3U^{-1}\right) \\
      &\quad-P^\dagger \bsigma_3U^{-1} + U\bsigma_3U^{-1}P^\dagger U^{-1}\\
   &= \bigl[ l(-P^\dagger\bsigma_3U^{-1}P^\dagger\bsigma_3U^\dagger
                +2EUU^\dagger) \\
    &\quad\phantom{\bigl[}-P^\dagger U^\dagger
       +U\bsigma_3U^{-1}P^\dagger\bsigma_3U^\dagger
     \qquad\bigr] {U^\dagger}^{-1}\bsigma_3U^{-1}
  }
\end{eqnarray}
Due to the constraint~(\ref{ConstHam}),
\begin{equation}
  P^\dagger\bsigma_3 U^\dagger = U \bsigma_3 P \;,
\end{equation}
so that equation~(\ref{epsPoisson}) simplifies to
\begin{eqnarray}
  \eqalign{
    \left\{\boldsymbol\epsilon,{\cal H}\right\} 
   &= \big[l(-P^\dagger P+2EU^\dagger U) - P^\dagger U^\dagger + U P\big]
      {U^\dagger}^{-1}\bsigma_3 U^{-1} \\
   &= \big[-2l{\cal H} + 2\proj{UP}_2\big]
      {U^\dagger}^{-1}\bsigma_3 U^{-1} \\
   &= 0 \;.
  }
\end{eqnarray}
Thus, the Lenz vector $\bepsilon$ is actually conserved.

\section{The Kustaanheimo-Stiefel description of closed orbits}
\label{sec:KSClosed}

If the initial conditions $\bi x(0)$ and $\dot{\bi x}(0)$ for a
trajectory are given, the pertinent initial conditions for the spinors $U$
and $U'$ can usually be obtained, up to a choice of gauge, from
(\ref{invKS}) and (\ref{KSderivTau}). This prescription fails for
trajectories starting at the origin, where the KS transformation is
singular. Although at first sight these trajectories appear rather
exceptional, they bear a particular significance to atomic physics:
Semiclassical closed-orbit theory \cite{Du88,Bogomolny89,Bartsch03b}
respresents an atomic photo-absorption spectrum as a sum over closed
classical orbits, i.e. orbits that start at the nucleus and return to it.
Due to that particular importance, closed orbits deserve a detailed
discussion. In this section, I will derive initial conditions for
a trajectory  starting at the Coulomb centre. I will then construct an
orthonormal basis of the KS spinor space that is suitable for an
investigation of the stability of orbits closed at the centre because it
allows to separate the physically distinct directions along and transverse
to the orbit from the gauge degree of freedom.

To find initial conditions for orbits starting at the nucleus note that in
the vicinity of the nucleus the Coulomb interaction is so strong that it
dominates all external forces. The dynamics close to the nucleus is
therefore described by the Kepler equation of motion~(\ref{UKepler}) and
its solution~(\ref{KeplerSol}). With the  initial condition $U(\tau=0)=0$
implemented, (\ref{KeplerSol}) reads
\begin{equation}
  \label{U0Sol}
  U(\tau) = \frac{U'_0}{\sqrt{-2E}}\,\sin(\sqrt{-2E}\,\tau) \;.
\end{equation}
$U'_0$ is the initial velocity in KS-coordinates. It must be
normalized to 
\begin{equation}
  \label{U0Norm}
  U_0^{\prime\dagger} U'_0 = 4 \;.
\end{equation}
The choice of gauge for $U'_0$ is arbitrary.

The position vector corresponding to (\ref{U0Sol}) is
\begin{equation}
  \bi x(\tau)=\frac12 U'_0\bsigma_3 U_0^{\prime\dagger}
               \,\frac{\sin^2(\sqrt{-2E}\,\tau)}{-2E} \;.
\end{equation}
Thus, (\ref{U0Sol}) describes an electron moving out from the nucleus in the
direction of the unit vector
\begin{equation}
  \label{U0Rot}
  \bi s = \frac14 U'_0\bsigma_3 U_0^{\prime\dagger} \;.
\end{equation}
$U_0'$ is therefore a spinor rotating the vector $\bsigma_3$ to the
starting direction $\bi s$ and normalized according to
(\ref{U0Norm}). In terms of the starting angles $\vartheta$ and $\varphi$
it is given by
\begin{equation}
  \label{ClosedStart}
  U_0'=2\,\rme^{-I_3\varphi/2}\rme^{-I_2\vartheta/2}\rme^{-I_3\alpha/2}
\end{equation}
with an arbitrary gauge parameter $\alpha$. The exponentials
in~(\ref{ClosedStart}) describe a sequence of three rotations taking the
reference vector~$\bsigma_3$ to the starting direction~$\bi s$.  The
initial momentum reads
\begin{equation}
  \label{ClosedStartP}
  P_0=U_0^{\prime\dagger}
     =2\,\rme^{I_3\alpha/2}\rme^{I_2\vartheta/2}\rme^{I_3\varphi/2} \;.
\end{equation}
Its component decomposition is
\begin{eqnarray}
  \label{ClosedStartPComp}
  \eqalign{
    p_0&=\phantom{-}2\cos\frac{\vartheta}{2}\cos\frac{\varphi+\alpha}{2} \;, \\
    p_1&=\phantom{-}2\sin\frac{\vartheta}{2}\sin\frac{\varphi-\alpha}{2} \;, \\
    p_2&=-2\sin\frac{\vartheta}{2}\cos\frac{\varphi-\alpha}{2} \;, \\
    p_3&=-2\cos\frac{\vartheta}{2}\sin\frac{\varphi+\alpha}{2} \;.
  }
\end{eqnarray}

To describe the stability of a classical trajectory, a coordinate system
with one coordinate along the trajectory and two coordinates perpendicular
to it is customarily introduced in the neighbourhood of the trajectory. A
linear stability analysis then requires calculating the derivatives of
positions and momenta with respect to the transverse initial
conditions. Most conveniently, derivatives with respect to two orthonormal
directions can be used. If these derivatives are to be calculated within
the framework of the KS theory, for a given starting direction $\bi s$ and
a direction $\bi s_\omega\bot\bi s$, a KS spinor $P_\omega$ must be found
such that a variation of the initial KS momentum $P_0$ in the direction of
$P_\omega$ corresponds to a variation of $\bi s$ in the direction of $\bi
s_\omega$.

As the initial momentum is given in terms of the starting angles
in~(\ref{ClosedStartP}), the derivatives $\partial P_0/\partial\vartheta$
and $\partial P_0/\partial\varphi$ can be expected to describe variations
of the starting direction in the directions of increasing $\vartheta$ and
$\varphi$, respectively. To check this and to find the correct
normalization of the spinors, I will now construct, for a fixed direction
$\bi s$, a basis of the spinor space such that one of the basis spinors
describes a variation of initial momentum along the orbit, i.e.~in the
direction of $\bi s$, two give variations in the directions of two
perpendicular unit vectors, and the fourth basis spinor describes the gauge
degree of freedom introduced by the KS regularization.

To this end, I consider a family of trajectories parameterized by an
arbitrary parameter $\omega$. All trajectories start at the nucleus, the
starting direction is given by a family of vectors $\bi s(\omega)$.
By~(\ref{U0Rot}), the initial KS momenta $P_0(\omega)$ then satisfy
\begin{equation}
  \bi s(\omega)=\frac14 P_0^\dagger(\omega) \bsigma_3 P_0(\omega) \;,
\end{equation}
so that in complete analogy with (\ref{KSderiv}) and (\ref{KSderiv2})
\begin{equation}
  \frac{\partial\bi s}{\partial\omega} = 
    \frac12 \left\langle P_0^\dagger\bsigma_3
                         \frac{\partial P_0}{\partial\omega}
            \right\rangle_1
\end{equation}
and
\begin{equation}
  \label{P0Var}
  \frac{\partial P_0}{\partial\omega}=
    \frac12 \bsigma_3 P_0 \frac{\partial\bi s}{\partial\omega}=
    \frac12 P_0\bi s \frac{\partial\bi s}{\partial\omega}
\end{equation}
if the gauge condition
\begin{equation}
  \label{P0gauge}
  \left\langle P_0^\dagger\bsigma_3\frac{\partial P_0}{\partial\omega}
  \right\rangle_3=0
\end{equation}
is imposed.
Equation (\ref{P0Var}) gives the variation in the initial KS momentum
pertinent to a given variation in the starting direction. If two different
variations are given, the scalar product of the momentum variations is
\begin{equation}
  \left(\frac{\partial P_0}{\partial\omega_1}\right)^\dagger \ast
  \frac{\partial P_0}{\partial\omega_2} = 
  \frac{\partial\bi s}{\partial\omega_1} \ast
  \frac{\partial\bi s}{\partial\omega_2} \;,
\end{equation}
so that the variations of KS momentum calculated from (\ref{P0Var}) are
orthonormal if the prescribed variations of the starting direction are.

For a fixed starting direction
\begin{equation}
  \bi s = \rme^{-I_3\varphi/2}\rme^{-I_2\vartheta}\bsigma_3
           \rme^{I_3\varphi/2}
\end{equation}
given by the starting angles $\vartheta$ and $\varphi$, I now introduce
the orthogonal vectors
\begin{eqnarray}
  \label{KSVarS}
  \eqalign{
  \bi s_{\vartheta} &= \rme^{-I_3\varphi/2}\rme^{-I_2\vartheta}\bsigma_1
                        \rme^{I_3\varphi/2} \;, \\
  \bi s_{\varphi} &= \rme^{-I_3\varphi/2}\bsigma_2
                      \rme^{I_3\varphi/2} \;.
  }
\end{eqnarray}
These are the unit vectors in the directions of $\partial\bi s/
\partial\vartheta$ and $\partial\bi s/\partial\varphi$, respectively. The
orthonormal basis $\bi s, \bi s_\vartheta$ and $\bi s_\varphi$ of the
position space gives rise to the three orthonormal KS spinors
\begin{equation}
  \begin{array}{r@{}l@{}l}
    P_s =& \frac12 \bsigma_3 P_0 \bi s &= \frac12 P_0 \;,\\[.5ex]
    P_\vartheta =& \frac12 \bsigma_3 P_0 \bi s_\vartheta
                &= \frac{I_2}{2} \rme^{-I_3\alpha}P_0 \;, \\[.5ex]
    P_\varphi   =& \frac12 \bsigma_3 P_0 \bi s_\varphi
                &= -\frac{I_1}{2} \rme^{-I_3\alpha}P_0 \;.
  \end{array}
\end{equation}
This set is complemented by a fourth orthonormal spinor 
\begin{equation}
  P_\alpha=\frac{\partial P_0}{\partial\alpha}=\frac 12\bsigma_3P_0 I \;.
\end{equation}
(Note the analogy with (\ref{P0Var}).) The spinor $P_\alpha$ maximally
violates the gauge condition (\ref{P0gauge}) in the sense that
\begin{equation}
  P_0^\dagger\bsigma_3P_\alpha =
    \proj{P_0^\dagger\bsigma_3P_\alpha}_3 \;.
\end{equation}
It therefore gives the direction in spinor space that corresponds to a gauge
transformation, whereas $P_s$ describes a change of momentum along the
orbit (i.e. a change of the energy) and $P_\vartheta$ and $P_\varphi$ give
directions perpendicular to the trajectory. The desired separation of the
physically distinct degrees of freedom has thus been achieved. Note that
$P_\vartheta=\partial P_0/\partial\vartheta$ as anticipated, whereas
$\partial P_0/\partial\varphi=P_0 I_3/2$ does not satisfy
(\ref{P0gauge}). Instead,
\begin{equation}
  \frac{\partial P_0}{\partial\varphi} =
    P_\alpha\cos\vartheta + P_\varphi \sin\vartheta \;.
\end{equation}

In components, the four basis spinors read
\begin{eqnarray}
  \label{PVarKomp}
  \fl\eqalign{
    P_s &= \textstyle\phantom{-}
	         \cos\frac{\vartheta}{2}\cos\frac{\varphi+\alpha}{2}
          -I_1 \sin\frac{\vartheta}{2}\sin\frac{\varphi-\alpha}{2} 
          +I_2 \sin\frac{\vartheta}{2}\cos\frac{\varphi-\alpha}{2}
          +I_3 \cos\frac{\vartheta}{2}\sin\frac{\varphi+\alpha}{2} \;,\\
    P_\vartheta &= \textstyle
	        -\sin\frac{\vartheta}{2}\cos\frac{\varphi+\alpha}{2}
          -I_1 \cos\frac{\vartheta}{2}\sin\frac{\varphi-\alpha}{2} 
          +I_2 \cos\frac{\vartheta}{2}\cos\frac{\varphi-\alpha}{2}
          -I_3 \sin\frac{\vartheta}{2}\sin\frac{\varphi+\alpha}{2} \;,\\
    P_\varphi &= \textstyle
	        -\sin\frac{\vartheta}{2}\sin\frac{\varphi+\alpha}{2}
          -I_1 \cos\frac{\vartheta}{2}\cos\frac{\varphi-\alpha}{2} 
          -I_2 \cos\frac{\vartheta}{2}\sin\frac{\varphi-\alpha}{2}
          +I_3 \sin\frac{\vartheta}{2}\cos\frac{\varphi+\alpha}{2} \;,\\
    P_\alpha &= \textstyle
	        -\cos\frac{\vartheta}{2}\sin\frac{\varphi+\alpha}{2}
          +I_1 \sin\frac{\vartheta}{2}\cos\frac{\varphi-\alpha}{2} 
          +I_2 \sin\frac{\vartheta}{2}\sin\frac{\varphi-\alpha}{2}
          +I_3 \cos\frac{\vartheta}{2}\cos\frac{\varphi+\alpha}{2} \;.\\
  }
\end{eqnarray}
These formulae prescribe a basis of the spinor space uniquely up to the
choice of $\alpha$ if $\vartheta\ne0,\pi$. At the poles, the angle
$\varphi$ is undefined. Because in this case (\ref{KSVarS}) gives a pair of
orthonormal tangent vectors for any choice of $\varphi$, (\ref{PVarKomp})
can be used with arbitrary $\varphi$.  The basis $P_s, P_\vartheta,
P_\varphi, P_\alpha$ of the KS momentum space can be supplemented by
position spinors $U_j=P_j^\dagger$ to obtain the basis of a canonical
coordinate system in spinor space. This basis set can then be used to
investigate the stability properties of closed orbits.

As a particular application, consider the stability determinant
\begin{equation}
  \label{MPrDef}
  M'=\det\frac{\partial(p_{\vartheta_f},p_{\varphi_f})}
              {\partial(\vartheta_i,\varphi_i)}
\end{equation}
occurring in the closed-orbit theory description of atoms in crossed-fields
\cite{Ellerbrock91,Main91,Bartsch02,Bartsch03b}. It encodes the stability
of a closed orbit starting from the nucleus in the direction characterized
by the angles $\vartheta_i, \varphi_i$ and returning from the direction
$\vartheta_f,\varphi_f$. As it stands, (\ref{MPrDef}) is not well suited to
practical computations because it suffers from the singularities of the
spherical coordinate chart: At the poles, neither the angle $\varphi$ nor
the angular momenta $p_\vartheta$ and $p_\varphi$ are well defined, so that
close to the poles, the calculation of the determinant becomes numerically
unstable. With the help of the spinor basis~(\ref{PVarKomp}),
(\ref{MPrDef}) can be rewritten in a form that is not plagued by any
singularities.

For a trajectory returning
to the nucleus at time~$\tau=0$ with a final KS momentum $P_f$, the
solution~(\ref{U0Sol}) of the Kepler equation of motion takes the form
\begin{equation}
  \label{KSReturn}
  \eqalign{
  U(\tau)=\frac{P_f^\dagger}{\sqrt{-2E}}\,\sin(\sqrt{-2E}\,\tau)
         =-\sqrt{\frac r2}P_f^\dagger\;, \\
  P(\tau)=P_f\cos(\sqrt{-2E}\,\tau)=\sqrt{1+Er}P_f \;,
  }
\end{equation}
which is valid as soon as the electron is sufficiently close to the nucleus
so that the external fields can be neglected. In~(\ref{KSReturn}), the
normalization condition $U^\dagger U = 2r$ and the pseudo-energy
conservation~(\ref{KepHam}) were used.

To discuss the stability properties of an orbit, consider a family of
orbits parameterized by an arbitrary parameter $\omega_1$. Equations of
motion for the derivatives $\partial U/\partial\omega_1$ and $\partial
P/\partial \omega_1$ are then obtained by linearizing the equations of
motion for $U$ and $P$ around the unperturbed trajectory.  Close to the
nucleus the dynamics in governed by the linear equations~(\ref{KepHamEq}),
so that the perturbations satisfy the same equations of motion as the
coordinates and momenta. In analogy with~(\ref{KeplerSol}), they read
\begin{equation}
  \label{KSVar1}
  \eqalign{
    \frac{\partial U}{\partial\omega_1} &=
     \frac{M_1^\dagger}{\sqrt{-2E}}\,\sin(\sqrt{-2E}\,\tau)
    +M_2 \cos(\sqrt{-2E}\,\tau) \;, \\
    \frac{\partial P}{\partial\omega_1} &=
     M_1 \cos(\sqrt{-2E}\,\tau)
    -\sqrt{-2E}\,M_2^\dagger \sin(\sqrt{-2E})
  }
\end{equation}
with constant even multivectors $M_1$ and $M_2$. With the help of the
normalizations inferred from~(\ref{KSReturn}), equation~(\ref{KSVar1})
simplifies to
\begin{equation}
  \label{KSVar}
  \eqalign{
    \frac{\partial U}{\partial\omega_1} &=
     -\sqrt{\frac r2}\,M_1^\dagger
     +\sqrt{1+Er}\,M_2 \;, \\
    \frac{\partial P}{\partial\omega_1} &=
     \sqrt{1+Er}\,M_1
     -\sqrt{2r}E\,M_2^\dagger
  \; }
\end{equation}
so that the constants $M_1=\partial P_f/\partial\omega_1$ and $M_1=\partial
U_f/\partial\omega_1$ can finally be identified with the values of the
derivatives obtained at $r=0$.

According to (\ref{lKS}), the angular momentum component in a plane
specified by a bivector $B$ is
\begin{equation}
  L_B = L\ast B = \frac 12 \proj{BUP} \;.
\end{equation}
The derivative of $L_B$ with respect to a parameter $\omega_1$ can then be
calculated with the help of~(\ref{KSVar}). It is given by
\begin{equation}
  \label{LBVar}
  \frac{\partial L_B}{\partial\omega_1} =
    \frac 12\left\langle P_f B \frac{\partial U_f}{\partial\omega_1}
            \right\rangle \;.
\end{equation}
It does not depend on $r$ because all components of the angular momentum
bivector are conserved in the Coulomb region.

For the angular momentum component $p_{\omega_2}$ conjugate to an angular
coordinate $\omega_2$, the relevant bivector is
\begin{equation}
  B=\bi s \frac{\partial\bi s}{\partial\omega_2} \;,
\end{equation}
so that
\begin{equation}
  \frac{\partial p_{\omega_2}}{\partial\omega_1} = 
   \frac 12\left\langle P_f \bi s \frac{\partial\bi s}{\partial\omega_2}\,
                     \frac{\partial U_f}{\partial\omega_1}
           \right\rangle =
   \left\langle P_{\omega_2}\, \frac{\partial U_f}{\partial\omega_1}
   \right\rangle \;,
\end{equation}
where $P_{\omega_2}$ denotes the basis spinor corresponding to $\omega_2$
by (\ref{P0Var}) with the final momentum $P_f$ used in place of $P_0$.

Consider coordinates $\bar\vartheta$ and $\bar\varphi$, such that
$\partial\bi s/\partial\bar\vartheta=\bi s_\vartheta$ and $\partial\bi
s/\partial\bar\varphi=\bi s_\varphi$ are unit vectors given by
(\ref{KSVarS}). The stability determinant $M'$ can  then
be rewritten as
\begin{equation}
  \fl
  \det\frac{\partial(p_{\vartheta_f},p_{\varphi_f})}
           {\partial(\vartheta_i,\varphi_i)}
 =\det\left(
       \begin{array}{cc}
         \frac{\partial p_{\bar\vartheta_f}}
              {\rule{0pt}{1.5ex}\partial\bar\vartheta_i} &
         \sin\vartheta_i\frac{\partial p_{\bar\vartheta_f}}
                             {\rule{0pt}{1.5ex}\partial\bar\varphi_i}
       \\[1ex]
         \sin\vartheta_f\frac{\partial p_{\bar\varphi_f}}
                             {\rule{0pt}{1.5ex}\partial\bar\vartheta_i} &
         \sin\vartheta_i \sin\vartheta_f
           \frac{\partial p_{\bar\varphi_f}}
                {\rule{0pt}{1.5ex}\partial\bar\varphi_i}
       \end{array}
      \right) = \sin\vartheta_i \sin\vartheta_f M
\end{equation}
with a $2\times 2$-determinant
\begin{equation}
  \label{MDef}
  M = \det\left(
        \begin{array}{cc}
         \left\langle P_\vartheta\,\frac{\partial U_f}
                            {\rule{0pt}{1.5ex}\partial\bar\vartheta_i}
         \right\rangle &
         \left\langle P_\vartheta\,\frac{\partial U_f}
                            {\rule{0pt}{1.5ex}\partial\bar\varphi_i}
         \right\rangle \\[1ex]
         \left\langle P_\varphi\,\frac{\partial U_f}
                            {\rule{0pt}{1.5ex}\partial\bar\vartheta_i}
         \right\rangle &
         \left\langle P_\varphi\,\frac{\partial U_f}
                            {\rule{0pt}{1.5ex}\partial\bar\varphi_i}
         \right\rangle
	\end{array}
      \right)
\end{equation}
free of any coordinate-induced singularities.  It is this form of the
stability determinant that was used in \cite{Bartsch03b}.

The derivatives $\partial U_f/\partial\omega$ needed in (\ref{MDef}) can be
calculated numerically by integrating the linearized equations of motion
\begin{eqnarray}
  \label{HamLin}
  \eqalign{
    \frac{d}{d\tau}\,\frac{\partial U}{\partial\omega} &= \phantom{-}
      \left(\frac{\partial U}{\partial\omega}\ast\partial_U\right)
        \partial_P {\cal H} +
      \left(\frac{\partial P}{\partial\omega}\ast\partial_P\right)
        \partial_P {\cal H} \;, \\
    \frac{d}{d\tau}\,\frac{\partial P}{\partial\omega} &= -
      \left(\frac{\partial U}{\partial\omega}\ast\partial_U\right)
        \partial_U {\cal H} -
      \left(\frac{\partial P}{\partial\omega}\ast\partial_P\right)
        \partial_U {\cal H}
  }
\end{eqnarray}
along the closed orbit with initial conditions
\begin{equation}
  \frac{\partial U}{\partial\omega}(0) = 0 \;, \qquad
  \frac{\partial P}{\partial\omega}(0) = P_\omega \;.
\end{equation}

\section{Conclusion}

The geometric-algebra formulation of the KS transformation endows the four
KS coordinates with a clear geometric interpretation. It was shown in this
paper that it also allows for an easy incorporation into a Lagrangian and
Hamiltonian formulation of the KS theory. At the same time, the known
KS Hamiltonian, that was restricted to homogeneous external fields, was
generalized to describe the KS motion in arbitrary static external
electromagnetic fields. This result is of interest beyond the realm of
atomic physics because it is equally applicable to an arbitrary
conservative non-electromagnetic force.

Closed orbits starting at and returning to the Coulomb centre require a
special treatment that takes the singularity of the KS coordinate frame into
account. The geometric algebra offers a convenient way to derive closed
orbit initial conditions and a basis of the spinor space that separates
physically distinct degrees of freedom.

The calculations carried out here amply illustrate the power of the
geometric algebra formulation of the KS theory. It can thus be expected to
form a convenient starting point for analytic calculations in classical
perturbation theory.

\appendix

\section{Introduction to geometric algebra}
\label{app:GA}

Geometric algebra is an algebraic system designed to represent the
geometric properties of Euclidean space in the most comprehensive and
systematic way possible. It was pioneered by Hermann Grassmann and William
Kingdon Clifford during the nineteenth century. From the 1960's on, David
Hestenes, with the aim of providing a universal mathematical framework for
theoretical physics, extended the algebraic techniques of Grassmann and
Clifford by a differential and integral calculus within the geometric
algebra, which he called geometric calculus \cite{Hestenes84}.

This appendix gives only a sketch of geometric algebra as far as it is
needed in the present work. A more extensive introduction, with an
extension to Minkowski spacetime, is contained in
\cite{Gull93,Lasenby93}. A thorough introduction to the geometric algebra
of Euclidean 3-space, with a detailed discussion of applications to
classical mechanics, can be found in \cite{Hestenes90}. A detailed
presentation of the mathematical properties of the geometric algebra is
given in \cite{Hestenes84}.

\subsection{The geometric algebra of Euclidean 3-space}
\label{sec:GA}

The orientation of two vectors $\bi a$ and $\bi b$ in space can be
characterized by the projection of one vector onto the other, which is
described by the scalar product $\bi a\cdot\bi b$, and the plane spanned by
$\bi a$ and $\bi b$, which is characterized by the vector product $\bi
a\times\bi b$. In the geometric algebra these complementary products $\bi
a\cdot\bi b$ and $\bi a\times\bi b$ are unified into a single ``geometric''
product $\bi a\bi b$. The construction of the geometric product starts by
picking a right-handed frame of orthonormal unit vectors $\bsigma_1$,
$\bsigma_2$ and $\bsigma_3$. For them, the existence of an
associative, but non-commutative geometric product satisfying
\begin{equation}
  \label{CliffordDef}
  \bsigma_i\bsigma_j + \bsigma_j\bsigma_i = 2\delta_{ij}
\end{equation}
is assumed. In addition, the geometric product is required to obey the
distributive law with respect to the usual addition of vectors. It follows
from (\ref{CliffordDef}) that $\bsigma_i^2=1$ is a scalar.  The reader may
notice that the defining relation~(\ref{CliffordDef}) is the same as obeyed
by the Pauli spin matrices. Indeed, these matrices generate a matrix
representation of the Clifford algebra of Euclidean 3-space. In the present
context, however, it is important to retain the interpretation of the
$\bsigma_i$ as ordinary vectors instead of regarding them as matrices. The
elements of the Clifford algebra are thus given a geometric interpretation,
as is indicated by the name ``geometric algebra'' introduced by Clifford
himself. It turns out that all calculations within geometric algebra can be
done without recourse to a matrix representation.

By virtue of the defining relation~(\ref{CliffordDef}), the geometric
product of two arbitrary vectors $\bi a=\sum_{i=1}^3 a_i\bsigma_i$ and
$\bi b=\sum_{i=1}^3 b_i\bsigma_i$ is
\begin{eqnarray}
  \label{ab1}
    \bi a\bi b =& a_1b_1 + a_2 b_2 + a_3 b_3 \\
      & (a_2b_3-a_3b_2)\bsigma_2\bsigma_3 + 
	(a_3b_1-a_1b_3)\bsigma_3\bsigma_1 +
	(a_1b_2-a_2b_1)\bsigma_1\bsigma_2\;. \nonumber
\end{eqnarray}
The scalar terms of this equation comprise the scalar product $\bi
a\cdot\bi b$. In addition, there are terms containing the product of two
orthogonal vectors. These terms are neither scalars nor vectors. They are
referred to as bivectors. As their coefficients are the components of the
vector cross product $\bi a\times\bi b$, bivectors should be interpreted
as describing an oriented area in the same way as a vector describes an
oriented line segment. Accordingly, a product of three orthonormal vectors
is called a trivector and interpreted as representing an oriented volume
element. With the help of the unit trivector
\begin{equation}
  \label{IDef}
  I = \bsigma_1\bsigma_2\bsigma_3 \;,
\end{equation}
equation~(\ref{ab1}) can be rewritten as
\begin{equation}
  \label{ProdDec}
  \bi a\bi b = \bi a\cdot\bi b + I \bi a\times\bi b \;,
\end{equation}
which achieves the desired unification of the scalar and vector
products. Notice that~(\ref{ProdDec}) contains a sum of quantities of
different types, a scalar and a bivector. This should be regarded as a
formal sum combining quantities of different types into a single object
with a scalar and a bivector part, in analogy to how a real and an
imaginary number are added to yield a complex number.

As the scalar product is symmetric in its factors whereas the vector
product is anti-symmetric, these products can be recovered from the
geometric product via
\begin{eqnarray}
  \eqalign{
  \bi a\cdot\bi b &= \frac 12(\bi a\bi b+\bi b\bi a)\;, \\
  \bi a\times\bi b &= \frac 1{2I}(\bi a\bi b-\bi b\bi a) \;.
  }
\end{eqnarray}
In particular, parallel vectors commute under the geometric product
whereas perpendicular vectors anti-commute, and any vector $\bi a$
satisfies $\bi a\bi a =\bi a\cdot\bi a$.

It is a crucial feature of the geometric algebra that it contains elements
of different grades, viz. scalars, vectors, bivectors, and trivectors. A
general element can be written as a sum of these pure-grade components and
is called a multivector. The pure-grade parts of any multivector can be
extracted by means of a grade projector. Let $\proj{A}_k$ be the grade-$k$
part of the multivector $A$. Due to its particular importance, the scalar
projector can be abbreviated as $\proj{A}=\proj{A}_0$, and the scalar
product of two multivectors $A$ and $B$ is defined by
\begin{equation}
  \label{ScalarDef}
  A\ast B = \proj{AB} \;.
\end{equation}
For vectors, this agrees with the scalar dot product. Any two multivectors
commute under the scalar product:
\begin{equation}
  A\ast B = B\ast A \;.
\end{equation}
In a term containing different kinds of products, the scalar product as
well as the vector cross product are understood to take precedence over the
geometric product. This convention has already been used
in~(\ref{ProdDec}).

A multivector which contains only parts of even grades, i.e., scalars and
bivectors, is referred to as an even multivector. The even multivectors
form a subalgebra of the full geometric algebra.

Finally, the reversion $A^\dagger$ of a multivector $A$ is obtained by
interchanging the order of vectors in any geometric product. Thus,
bivectors and trivectors change sign under reversion, whereas scalars and
vectors remain unchanged. Formally, the reversion can be defined by the
properties $\bi a^\dagger=\bi a$ for any vector $\bi a$ and
\begin{equation}
    (AB)^\dagger = B^\dagger A^\dagger \;, \qquad\qquad
    (A+B)^\dagger = A^\dagger + B^\dagger
\end{equation}
for multivectors $A$ and $B$.

\subsection{Rotations in the geometric algebra}
\label{sec:GARot}

Within the geometric algebra rotations are conveniently represented in the
form
\begin{equation}
  \label{RotGen}
  \bi a \mapsto {\rm R}(\bi a) = R\bi a R^\dagger
\end{equation}
with an even multivector $R$ satisfying the normalization
condition
\begin{equation}
  RR^\dagger=1 \;.
\end{equation}
Conversely, any normalized even multivector describes a rotation.
An arbitrary even multivector satisfies $\alpha=UU^\dagger \ge 0$, so that
$U=\sqrt{\alpha}R$ is a multiple of a rotor $R$. Therefore,
\begin{equation}
  U\bi aU^\dagger = \alpha R\bi aR^\dagger \;,
\end{equation}
and $U$ describes a rotation-dilatation of 3-space. In particular,
\begin{equation}
  \label{UVector}
  U\bi aU^\dagger=\proj{U\bi aU^\dagger}_1
\end{equation}
is a vector for any even multivector $U$ and any vector $\bi a$.

A rotation can be characterized by specifying two vectors $\bi a$ and $\bi
b$ so that $\bi a$ is mapped to $\bi b$ by a rotation in the plane
$\proj{\bi a\bi b}_2$ spanned by $\bi a$ and $\bi b$. The rotor $R$
describing this rotation is
\begin{equation}
  \label{abRotor}
  R = \frac{1+\bi b\bi a}{|\bi a + \bi b|}
    = \frac{1+\bi b\bi a}{\sqrt{2(1+\bi a\cdot\bi b)}} \;.
\end{equation}
If, alternatively, a rotation is characterized by its rotation axis,
given by a unit vector $\bi n$, and a rotation angle $\varphi$, the
pertinent rotor reads
\begin{equation}
  R = \rme^{-I\bi n\varphi/2} \;,
\end{equation}
where the exponential function of an arbitrary multivector is defined by
the familiar power series
\begin{equation}
  \label{ExpDef}
  \rme^A = \sum_{n=0}^\infty \frac{A^n}{n!} \;.
\end{equation}
It satisfies the ``power law'' relation
\begin{equation}
  \label{PowerLaw}
  \rme^{A+B} = \rme^A \rme^B
\end{equation}
if $AB=BA$ and
\begin{eqnarray}
  \label{ExpCommut}
  \rme^{A}B = B\rme^{A}  &\qquad\textrm{if } AB=BA\;, \\
  \rme^{A}B = B\rme^{-A} &\qquad\textrm{if } AB=-BA\;.
\end{eqnarray}

\subsection{The multivector derivative}
\label{sec:multDeriv}

The formalism of the multivector derivative provides a differential
calculus for arbitrary multivector functions. Let $F(X)$ be a smooth
multivector-valued function of the multivector argument $X$. Neither the
grades contained in $X$ nor in $F$ are specified. The
directional derivative in the direction of a fixed multivector $A$ by
\begin{equation}
  \label{DirectDeriv}
  A\ast \partial_X F(X) = \left.\frac{dF(X+\tau P_X(A))}{d\tau}\right|_
                           {\tau=0}  \;,
\end{equation}
where $P_X(A)$ projects $A$ onto the grades contained in
$X$. (\ref{DirectDeriv}) agrees with the familiar definition of the
directional derivative.

Let $e_J, J=1,\dots,8$ be a basis of the geometric algebra and $e^J$ its
dual basis, i.e., $e_J\ast e^K = \delta_J^K$. The multivector derivative is
then defined to be
\begin{equation}
  \label{MultDeriv}
  \partial_X = \sum_J e^J \, e_J\ast\partial_X \;.
\end{equation}
It inherits the algebraic properties of its argument $X$. In particular,
$\partial_X$ contains the same grades as $X$. Notice that the scalar
product $A\ast\partial_X$ is indeed the directional derivative in the
direction $A$, justifying the notation introduced
in~(\ref{DirectDeriv}). For a vector argument $\bi x$, the multivector
derivative $\partial_{\bi x}$ reduces to the vector derivative, which is
analogous to the familiar nabla operator.

Both the directional derivative and the multivector derivative are linear
operators and satisfy Leibniz' rule
\begin{eqnarray}
  \label{prodDirect}
  \fl
  A\ast\partial_X \big(F(X)G(X)\big) = 
     \big(A\ast\partial_X F(X)\big) G(X) 
   + F(X) \big(A\ast\partial_X G(X)\big) \;, \\
  \label{prodMult}
  \fl
  \partial_X(F(X)G(X)) 
    = \stackrel{*}{\partial}_X\stackrel{*}{F}(X)G(X)
     +\stackrel{*}{\partial}_XF(X)\stackrel{*}{G}(X) \;.
\end{eqnarray}
In~(\ref{prodMult}), the overstars indicate the functions to be
differentiated. Notice that the second term in~(\ref{prodMult}) is in
general different from $F(X)\big(\partial_X G(X)\big)$, because due to its
multivector properties the multivector derivative does not commute with $F$
even if $F(X)$ is not differentiated. The directional derivative, on the
contrary, is a scalar differential operator that commutes with any
multivector that is not to be differentiated. For this reason it is often
convenient to write the multivector derivative as
\begin{equation}
  \label{MultDirect}
  \partial_X = \partial_A\,A\ast\partial_X \;.
\end{equation}
This form decomposes $\partial_X$ into a multivector $\partial_A$ and a
scalar differential operator $A\ast\partial_X$, which can be moved freely
among multivectors.

In addition, the directional derivative satisfies the chain rule
\begin{equation}
  \label{chainDirect}
  A\ast\partial_X F(G(X)) =
      \big(A\ast\partial_X G(X)\big)\ast\partial_G F(G) \;,
\end{equation}
which is useful in many calculations.

A fundamental result concerning the multivector derivative is
\begin{equation}
  \label{MultProject}
  \partial_X\proj{XA} = \partial_X\proj{AX} = P_X(A)
\end{equation}
for any multivector $A$. As a consequence,
\begin{equation}
  \partial_X\proj{X^\dagger A} = \partial_X\proj{AX^\dagger} =
  P_X(A^\dagger) \;.
\end{equation}

\section*{References}

\end{document}